\documentclass[twocolumn,showpacs,preprintnumbers]{revtex4}
\usepackage{amsfonts}
\usepackage{amsmath}
\usepackage{graphicx}
\usepackage{dcolumn}
\usepackage{bm}
\usepackage{mathrsfs}
\usepackage{epsfig}

\begin{document}

\preprint{APS/123-QED}
\title{Kinetic axi-symmetric gravitational equilibria\\
in collisionless accretion disc plasmas}
\author{Claudio Cremaschini}
\email{cremasch@sissa.it}
\author{John C. Miller}
\altaffiliation[Also at ]{Department of Physics (Astrophysics), Oxford University, Oxford, U.K.}
\affiliation{International School for Advanced Studies (SISSA) and INFN, Trieste, Italy}
\author{Massimo Tessarotto}
\altaffiliation[Also at ]{Consortium for Magnetofluid Dynamics, Trieste University, Trieste, Italy}
\affiliation{Department of Mathematics and Informatics, Trieste University, Trieste, Italy}
\date{\today }

\begin{abstract}
A theoretical treatment is presented of kinetic equilibria in accretion
discs around compact objects, for cases where the plasma can be considered
as collisionless. The plasma is assumed to be axi-symmetric and to be acted
on by gravitational and electromagnetic fields; in this paper, the
particular case is considered where the magnetic field admits a family of
toroidal magnetic surfaces, which are locally mutually-nested and closed. It
is pointed out that there exist asymptotic kinetic equilibria represented by
generalized bi-Maxwellian distribution functions and characterized by
primarily toroidal differential rotation and temperature anisotropy. It is
conjectured that kinetic equilibria of this type can exist which are able to
sustain both toroidal and poloidal electric current densities, the latter
being produced via finite Larmor-radius effects associated with the
temperature anisotropy. This leads to the possibility of existence of a new
kinetic effect - referred to here as a \textquotedblleft kinetic dynamo
effect\textquotedblright\ - resulting in the self-generation of toroidal
magnetic field even by a stationary plasma, without any net radial accretion
flow being required. The conditions for these equilibria to occur, their
basic theoretical features and their physical properties are all discussed
in detail.
\end{abstract}

\maketitle








\section{Introduction}

This paper is concerned with dynamical processes in astrophysical accretion
disc (AD) plasmas and their relationship with the accretion process. The aim
of the research programme of which the first part is reported in the present
paper is to provide a consistent theoretical formulation of kinetic theory
for AD plasmas, which can then be used for investigating their equilibrium
properties and dynamical evolution. Note that what is meant here by the term
\textquotedblleft equilibrium\textquotedblright\ is in general a
stationary-flow solution, which can also include a radial accretion
velocity. Apart from the intrinsic interest of this study for the
equilibrium properties of accretion discs, the conclusions reached may have
important consequences for other applications and for stability analyses of
the discs. In this paper we consider the particular case where the AD plasma
contains domains of locally-closed magnetic surfaces where there is in fact
no local net accretion. We then focus on these domains. We do this, both
because this represents a situation which is of considerable potential
interest, but also because it leads to some significant simplifications of
the discussion. In subsequent papers, we will proceed to consider more
general cases.

\subsection{Astrophysical background}

Accretion discs are observed in a wide range of astrophysical contexts, from
the small-scale regions around proto-stars or stars in binary systems to the
much larger scales associated with the cores of galaxies and Active Galactic
Nuclei (AGN). Observations tell us that these systems contain matter
accreting onto a central object, losing angular momentum and releasing
gravitational binding energy. This can give rise to an extremely powerful
source of energy generation, causing the matter to be in the plasma state
and allowing the discs to be detected through their radiation emission \cite%
{Frank, Vietri, Balbus1998}. A particularly interesting class of accretion
discs consists of those occurring around black holes in binary systems,
which give rise to compact X-ray sources. For these, one has both a strong
gravitational field and also presence of significant magnetic fields which
are mainly self-generated by the plasma current densities. Despite the
information available about these systems, mainly provided by observations
collected over the past forty years and concerning their macroscopic
physical and geometrical properties (structure, emission spectrum, etc.), no
complete theoretical description of the physical processes involved in the
generation and evolution of the magnetic fields is yet available. While it
is widely thought that the magneto-rotational instability (MRI) \cite%
{Balbus1998} plays a leading role in generating an effective viscosity in
these discs, more remains to be done in order to obtain a full understanding
of the dynamics of disc plasmas and the relation of this with the accretion
process. This requires identifying the microphysical phenomena involved in
the generation of instabilities and/or turbulence which may represent a
plausible source for the effective viscosity which, in turn, is then related
to the accretion rates \cite{Frank, Balbus1998}. There is a lot of
observational evidence which cannot yet be explained or fully understood
within the framework of existing theoretical descriptions and many
fundamental questions still remain to be answered \cite{Sera01, Melia}.

\bigskip

\subsection{Motivations for a kinetic theory}

Historically, most theoretical and numerical investigations of accretion
discs have been made in the context of hydrodynamics (HD) or
magneto-hydro-dynamics (MHD) (\textit{fluid approaches}) \cite%
{Ferraro1937,Mestel1961,Lovelace1976,Blanford1976,Blanford-Payne1982,Frank,
Vietri, Balbus1998,Miller2001}. Treating the medium as a fluid allows one to
capture the basic large-scale properties of the disc structure and
evolution. An interesting development within this context has been the work
of Coppi \cite{Coppi1} and Coppi and Rousseau \cite{Coppi2} who showed that
stationary magnetic configurations in AD plasmas, for both low and high
magnetic energy densities, can exhibit complex magnetic structures
characterized locally by plasma rings with closed nested magnetic surfaces.
However, even the most sophisticated fluid models are still not able to give
a good explanation for all of the complexity of the phenomena arising in
these systems. While fluid descriptions are often useful, it is well known
that, at a fundamental level, a correct description of microscopic and
macroscopic plasma dynamics should be formulated on the basis of kinetic
theory (\textit{kinetic approach}) \cite{Cremasch2008-1, Cremasch2008-2},
and for that there is still, remarkably, no satisfactory theoretical
formulation. Going to a kinetic approach can overcome the problem
characteristic of fluid theories of uniquely defining consistent closure
conditions \cite{Cremasch2008-1, Cremasch2008-2}, and a kinetic formulation
is necessarily required, instead of MHD, for correctly describing regimes in
which the plasma is either collisionless or weakly collisional \cite%
{Quataert2002, Quataert2007, Quataert2007B}. In these situations, the
distribution function describing the AD plasma will be different from a
Maxwellian, which is instead characteristic of highly collisional plasmas
for which fluid theories properly apply. An interesting example of
collisionless plasmas, arising in the context of astrophysical accretion
discs around black holes, is that of radiatively inefficient accretion flows 
\cite{NarayanA,NarayanB}. Theoretical investigations of such systems have
suggested that the accreting matter consists of a two-temperature plasma,
with the proton temperature being much higher than the electron temperature 
\cite{Sera01, Melia}. This in turn implies that the timescale for energy
exchange by Coulomb collisions between electrons and ions must be longer
than the other characteristic timescales of the system, in particular the
inflow time. In this case, a correct physical description of the
phenomenology governing these objects can only be provided by a kinetic
formulation. Finally, the kinetic formalism is more convenient for the
inclusion of some particular physical effects, including ones due to
temperature anisotropies, and is essential for making a complete study of
the kinetic instabilities which can play a key role in causing the accretion
process \cite{Quataert2002, Quataert2007, Quataert2007B}. We note here that,
although there are in principle several possible physical processes which
may explain the appearance of temperature anisotropies (see for example \cite%
{Quataert2007, Quataert2007B}), the main reason for their maintenance in a
collisionless and non-turbulent plasma may simply be the lack of any
efficient mechanism for temperature isotropization.

\bigskip

\subsection{Previous work}

Only a few studies have so far addressed the problem of deriving a kinetic
formulation of steady-state solutions for AD plasmas.

The paper by Bhaskaran and Krishan \cite{Mahajan01}, based on theoretical
results obtained by Mahajan \cite{Mahajan02, Mahajan03} for laboratory
plasmas, is a first example going in this direction. These authors assumed
an equilibrium distribution function expressed as an infinite power series
in the ratio of the drift velocity to the thermal speed (considered as the
small expansion parameter) such that the zero-order term coincides with a
homogeneous Maxwellian distribution. Assuming a prescribed profile for the
external magnetic field and ignoring the self-generated field, they looked
for analytic solutions of the Vlasov-Maxwell system for the coefficients of
the series (typically truncated after the first few orders). However, the
assumptions made strongly limited the applicability of their model.

Another approach proposed recently by Cremaschini \textit{et al.} \cite%
{Cremasch2008-2}, gives an exact solution for the equilibrium Kinetic
Distribution Function (KDF) of strongly magnetized non-relativistic
collisionless plasmas with isotropic temperature and purely toroidal flow
velocity. The strategy adopted was similar to that developed by Catto 
\textit{et al.} \cite{Catto1987} for toroidal plasmas, suitably adapted to
the context of accretion discs. The stationary KDF was expressed in terms of
the first integrals of motion of the system showing, for example, that the
standard Maxwellian KDF is an asymptotic stationary solution only in the
limit of a strongly magnetized plasma, and that the spatial profiles of the
fluid fields are fixed by specific kinetic constraints \cite{Cremasch2008-2}.

In recent years, the kinetic formalism has also been used for investigating
stability of AD plasmas, particularly in the collisionless regime and
focused on studying the role and importance of MRI \cite{Balbus1998,
Quataert2007B, Balbus1991, Snyder1997, Sharma2006}. The main goal of these
studies \cite{Quataert2002, Quataert2007, Quataert2007B, Snyder1997} was to
provide and test suitable kinetic closure conditions for
asymptotically-reduced fluid equations (referred to as \textquotedblleft
kinetic MHD\textquotedblright ), so as to allow the fluid stability analysis
to include some of the relevant kinetic effects for collisionless plasmas 
\cite{Quataert2002, Quataert2007B, Snyder1997}. However, none of them
systematically treated the issue of kinetic equilibrium, and the underlying
unperturbed plasma was usually taken to be described by either a Maxwellian
or a bi-Maxwellian KDF.

Finally, increasing attention has been paid to the role of temperature
anisotropy and the related kinetic instabilities. Some recent numerical
studies \cite{Sharma2006, Quataert2007B} have tried to include the effects
of temperature anisotropy but, although it is clear that this can give rise
to an entirely new class of phenomena, all of these estimates rely on fluid
models in which kinetic effects are included in only an approximate way. A
kinetic approach is needed rather than a fluid one, in order to give a clear
and self-consistent picture, and this needs to be based on equilibrium
solutions suitable for accretion discs.

\bigskip

\subsection{Open problems}

Many problems remain to be addressed and solved regarding the kinetic
formulation of AD plasma dynamics. Among them, we focus on the following:

\begin{enumerate}
\item The construction of a kinetic theory for AD plasmas within the
framework of the Vlasov-Maxwell description, and the investigation of their
kinetic equilibrium properties.

\item The inclusion of finite Larmor-radius (FLR) effects in the MHD
equations. For magnetized plasmas, this can be achieved by making a kinetic
treatment and representing the KDF in terms of gyrokinetic variables
(Bernstein and Catto \cite{Catto1977,Catto1978,Bernstein1985,Bernstein1986}%
). The gyrokinetic formalism provides a simplified description of the
dynamics of charged particles in the presence of magnetic fields, thanks to
the symmetry of the Larmor gyratory motion of the particles around the
magnetic field lines. Therefore \emph{kinetic} and \emph{gyrokinetic theory}
are both fundamental tools for treating FLR effects in a consistent way.

\item The determination of suitable kinetic closure conditions to be used in
the fluid description of the discs. These should include the kinetic effects
of the plasma dynamics in a consistent way.

\item Extension of the known solutions to more general contexts, with the
inclusion of important effects such as temperature anisotropy.

\item Development of a kinetic theory for stability analysis of AD plasmas.
As already mentioned, this could throw further light on the physical
mechanism giving rise to the effective viscosity and the related accretion
processes. This is particularly interesting for collisionless plasmas with
temperature anisotropy, since only kinetic theory could be able to explain
how instabilities can originate and grow to restore the isotropic properties
of the plasma.
\end{enumerate}

\subsection{Goals of the paper}

The aim of this paper is to extend the investigation of kinetic equilibria
developed in an earlier paper \cite{Cremasch2008-2} to a more general class
of solutions. In particular, we pose here the problem of constructing
analytic solutions for \textit{exact kinetic and gyrokinetic axi-symmetric
gravitational equilibria}\ (see definition below) in accretion discs around
compact objects. The solution presented is applicable to collisionless
magnetized plasmas with temperature anisotropy and mainly toroidal flow
velocity. The kinetic treatment of the gravitational equilibria necessarily
requires that the KDF is itself a stationary solution of the relevant
kinetic equations. Ignoring possible weakly-dissipative effects, we shall
assume - in particular - that the KDF and the electromagnetic (EM) fields
associated with the plasma obey the system of Vlasov-Maxwell equations. The
only restriction on the form of the KDF, besides assuming its strict
positivity and it being suitably smooth in the relevant phase-space, is due
to the requirement that it must be a function only of the independent first
integrals of the motion or the adiabatic invariants for the system.

The paper is organized as follows. In Section 2 we discuss the conditions
for the existence of a kinetic equilibrium and we introduce the basic
assumptions for the formulation of the kinetic theory. Section 3 deals with
the first integrals of motion and the gyrokinetic adiabatic invariants of
the system. In Section 4 we construct the equilibrium KDF for AD plasmas
with non-isotropic temperature, giving also a useful asymptotic expansion
for this in the limit of strong magnetic fields. Section 5 deals with
calculation of the fluid moments of the stationary KDF. The limit of
isotropic temperature is then investigated in Section 6, while Section 7 is
devoted to analyzing the Maxwell equations. Finally, Section 8 contains
conclusions and a summary of the main results.

\section{Kinetic theory for accretion disc plasmas:\ basic assumptions}

We first discuss what is meant by asymptotic kinetic equilibria in the
present study and what are the physical conditions under which they can be
realized.

An asymptotic kinetic equilibrium must be one obtained within the context of
kinetic theory and must be described by the stationary Vlasov-Maxwell
equations. This means that the generic plasma KDF, $f_{s}$, must be a
solution of the stationary Vlasov equation, as will be the case if $f_{s}$
is expressed in terms of exact first integrals of the motion or adiabatic
invariants of the system, which in turn implies that $f_{s}$ for each
species must be an exact first integral of the motion or an adiabatic
invariant. The stationarity condition means that the equilibrium KDF cannot
depend explicitly on time, although in principle it could contain an
implicit time dependence via its fluid moments (in which case the kinetic
equilibrium does not correspond to a fluid equilibrium and there are
non-stationary fluid fields).

In the following we shall take the AD plasma to be: a) \emph{non-relativistic%
}, in the sense that it has non--relativistic species flow velocities, that
the gravitational field can be treated within the classical Newtonian
theory, and that the non-relativistic Vlasov kinetic equation is used as the
dynamical equation for the KDF;\ b) \emph{collisionless}, so that the mean
free path of the plasma particles is much longer than the largest
characteristic scale length of the plasma; c)\ \emph{axi-symmetric}, so that
the relevant dynamical variables characterizing the plasma (e.g., the fluid
fields) are independent of the toroidal angle $\varphi ,$ when referred to a
set of cylindrical coordinates $(R,\varphi ,z)$; d) acted on by both
gravitational and EM fields.

Also, we will focus on the situation where the equilibrium magnetic field $%
\mathbf{B}$ admits, at least locally, a family of nested axi-symmetric
closed toroidal magnetic surfaces $\left\{ \psi (\mathbf{r})\right\} \equiv
\left\{ \psi (\mathbf{r})=const.\right\} $, where $\psi $ denotes the
poloidal magnetic flux of $\mathbf{B}$ (see \cite{Coppi1, Coppi2} for a
proof of the possible existence of such configurations in the context of
astrophysical accretion discs; see also \cite{Cremasch2008-1, Cremasch2008-2}
for further discussions in this regard and Fig.1 for a schematic view of
such a configuration). In this situation, a set of magnetic coordinates ($%
\psi ,\varphi ,\vartheta $) can be defined locally, where $\vartheta $ is a
curvilinear angle-like coordinate on the magnetic surfaces $\psi (\mathbf{r}%
)=const.$ Each relevant physical quantity $A(\mathbf{r})$ can then be
expressed as a function of these magnetic coordinates, i.e. $A(\mathbf{r}%
)=A\left( \psi ,\vartheta \right) ,$ where the $\varphi $ dependence has
been suppressed due to the axi-symmetry. It follows that it is always
possible to write the following decomposition: $A=A^{\sim }+\left\langle
A\right\rangle ,$ where the \textit{oscillatory part }$A^{\sim }\equiv
A-\left\langle A\right\rangle $ contains the $\vartheta $-dependencies and $%
\left\langle A\right\rangle $ is the $\psi -$surface average of the function 
$A(\mathbf{r})$ defined on a flux surface $\psi (\mathbf{r})=const.$ as $%
\left\langle A\right\rangle =\xi ^{-1}\oint d\vartheta A(\mathbf{r}%
)/\left\vert \mathbf{B}\cdot \nabla \vartheta \right\vert ,$ with $\xi $
denoting $\xi \equiv \oint d\vartheta /\left\vert \mathbf{B}\cdot \nabla
\vartheta \right\vert $.

For definiteness, we shall consider here a plasma consisting of at least two
species of charged particles: one species of ions and one of electrons.

We also introduce some convenient dimensionless parameters which will be
used in constructing asymptotic orderings for the relevant quantities of the
theory. The first one, which enters into the construction of the gyrokinetic
theory, is defined as $\varepsilon _{M}\equiv \max \left\{ \frac{r_{Ls}}{L}%
,s=i,e\right\} $, where $r_{Ls}=v_{\perp ths}/\Omega _{cs}$ is the species
average Larmor radius, with $v_{\perp ths}=\left\{ T_{\perp s}/M_{s}\right\}
^{1/2}$ denoting the species thermal velocity perpendicular to the magnetic
field direction and $\Omega _{cs}=Z_{s}eB/M_{s}c$ denoting the species
Larmor frequency. Here $L$ is the characteristic length-scale of the
inhomogeneities of the EM field, defined as $L\sim L_{B}\sim L_{E}$, where $%
L_{B}$ and $L_{E}$ are the characteristic lengths of the gradients of the
absolute values of the magnetic field $\mathbf{B}\left( \mathbf{r},t\right) $
and the electric field $\mathbf{E}\left( \mathbf{r},t\right) $, defined as $%
\frac{1}{L_{B}}\equiv \max \left\{ \left\vert \frac{\partial }{\partial r_{i}%
}\ln B\right\vert ,i=1,3\right\} $ and $\frac{1}{L_{E}}\equiv \max \left\{
\left\vert \frac{\partial }{\partial r_{i}}\ln E\right\vert ,i=1,3\right\} $%
. For typical temperatures and magnetic fields in AD plasmas, $0<\varepsilon
_{M}\ll 1$.

The second parameter is the \textit{inverse aspect ratio} defined as $\delta
\equiv \frac{r_{\max }}{R_{0}}$, where $R_{0}$ is the radial distance from
the vertical axis to the center of the nested magnetic surfaces and $r_{\max
}$ is the average cross-sectional poloidal radius of the largest closed
toroidal magnetic surface; see Fig.1 for a schematic view of the
configuration geometry and the meaning of the notation introduced here.
Then, we impose the requirement $0<\delta \ll 1$, which is referred to as
\textquotedblleft small inverse aspect ratio ordering\textquotedblright .
The main motivation for introducing this ordering is that we are discussing
only local solutions where this asymptotic condition holds; this property
also follows from the results presented in \cite{Coppi1, Coppi2}, and has
aready been used in other previous work on the subject \cite{Cremasch2008-1,
Cremasch2008-2}. The requirement $\delta \ll 1$ is also needed in order to
satisfy the constraint condition imposed by Ampere's law, as discussed in
Sec. VII. We stress that the $\delta -$ordering here introduced is
consistent with the assumption of nested and closed magnetic surfaces that
are assumed to be localized in space.

Finally we introduce a parameter $\delta _{Ts}$ which measures the magnitude
of the species temperature anisotropy and is defined as $\delta _{Ts}\equiv 
\frac{T_{\parallel s}-T_{\perp s}}{T_{\parallel s}}$, where $T_{\parallel s}$
and $T_{\perp s}$ denote the parallel and perpendicular temperatures, as
measured with respect to the magnetic field direction.

Note that, in the following, we will use a prime \textquotedblleft\ $%
^{\prime }$ \textquotedblright\ to denote a dynamical variable defined at
the guiding-center position.

\begin{figure}[tbp]
\centering
\includegraphics[width=3.3in,height=2.3in]{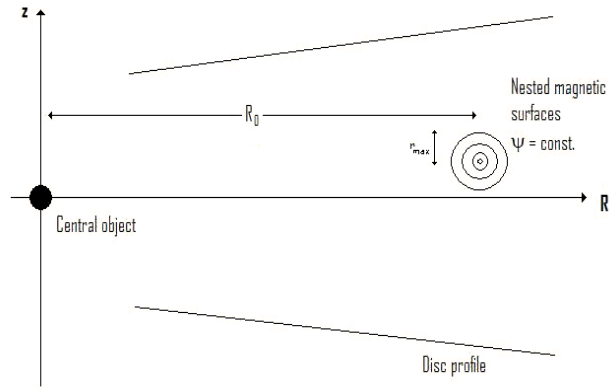} 
\caption{Schematic view of the configuration geometry.}
\end{figure}

\bigskip


\subsection{The treatment of EM and gravitational fields}

In the following, we shall assume that the magnetic field is of the form 
\begin{equation}
\left. \mathbf{B}\equiv \nabla \times \mathbf{A}=\mathbf{B}^{self}(\mathbf{r}%
,t)+\mathbf{B}^{ext}(\mathbf{r},t),\right.  \label{MAGNETIC FIELD}
\end{equation}%
where $\mathbf{B}^{self}$ and $\mathbf{B}^{ext}$ denote the self-generated
magnetic field produced by the AD plasma and a non-vanishing external
magnetic field produced by the central object. We also impose the following
relative ordering between the two components of the total magnetic field: $%
\frac{\left\vert \mathbf{B}^{ext}\right\vert }{\left\vert \mathbf{B}%
^{self}\right\vert }\sim O\left( \varepsilon _{M}^{k}\right) $, with $k\geq
1 $. This means that the self-field is the dominant component: the magnetic
field is primarily self-generated. Also, the overall magnetic field is
assumed to be slowly varying in time, i.e., to be of the form $\mathbf{B}(%
\mathbf{r},\varepsilon _{M}t)$, while $\mathbf{B}^{self}$ and $\mathbf{B}%
^{ext}$ are defined as 
\begin{eqnarray}
&&\left. \mathbf{B}^{self}=I(\mathbf{r},\varepsilon _{M}t)\nabla \varphi
+\nabla \psi _{p}(\mathbf{r},\varepsilon _{M}t)\times \nabla \varphi ,\right.
\label{POs-1} \\
&&\left. \mathbf{B}^{ext}=\nabla \psi _{D}(\mathbf{r},\varepsilon
_{M}t)\times \nabla \varphi ,\right.  \label{POS-2}
\end{eqnarray}%
where $\mathbf{B}_{T}\equiv I(\mathbf{r},\varepsilon _{M}t)\nabla \varphi $
and $\mathbf{B}_{P}\equiv \nabla \psi _{p}(\mathbf{r},\varepsilon
_{M}t)\times \nabla \varphi $ are the toroidal and poloidal components of
the self-field, with $(\psi _{p},\varphi ,\vartheta )$ defining locally a
set of magnetic coordinates. Moreover, the external magnetic field $\mathbf{B%
}^{ext}$ is assumed to be purely poloidal and defined in terms of the vacuum
potential $\psi _{D}(\mathbf{r},\varepsilon _{M}t)$. In particular, we
notice here that for typical astrophysical applications of interest, the
function $\psi _{D}(\mathbf{r},\varepsilon _{M}t)$ can be conveniently
identified with the flux function of a dipolar magnetic field. It follows
that the magnetic field can also be written in the form 
\begin{equation}
\mathbf{B}=I(\mathbf{r},\varepsilon _{M}t)\nabla \varphi +\nabla \psi (%
\mathbf{r},\varepsilon _{M}t)\times \nabla \varphi ,  \label{B FIELD}
\end{equation}%
where the function $\psi (\mathbf{r},\varepsilon _{M}t)$ is defined as $\psi
(\mathbf{r},\varepsilon _{M}t)\equiv \psi _{p}(\mathbf{r},\varepsilon
_{M}t)+\psi _{D}(\mathbf{r},\varepsilon _{M}t),$ and $(\psi ,\varphi
,\vartheta )$ define a set of local magnetic coordinates, as implied by the
equation $\mathbf{B}\cdot \nabla \psi =0$ which is identically satisfied. In
addition, it is assumed that the charged particles of the plasma are subject
to the action of \emph{effective} \emph{EM potentials} $\left\{ \Phi
_{s}^{eff}(\mathbf{r},\varepsilon _{M}t),\mathbf{A}(\mathbf{r},\varepsilon
_{M}t)\right\} ,$ where $\mathbf{A}(\mathbf{r},\varepsilon _{M}t)$ is the
vector potential corresponding to the magnetic field of Eq.(\ref{B FIELD}),
while $\Phi _{s}^{eff}(\mathbf{r},\varepsilon _{M}t)$ is given by%
\begin{equation}
\Phi _{s}^{eff}(\mathbf{r},\varepsilon _{M}t)=\Phi (\mathbf{r},\varepsilon
_{M}t)+\frac{M_{s}}{Z_{s}e}\Phi _{G}(\mathbf{r},\varepsilon _{M}t),
\label{potenziale efficace}
\end{equation}%
with $\Phi _{s}^{eff}(\mathbf{r},\varepsilon _{M}t),$ $\Phi (\mathbf{r}%
,\varepsilon _{M}t)$ and $\Phi _{G}(\mathbf{r},\varepsilon _{M}t)$ denoting
the \emph{effective} electrostatic potential and the electrostatic and
generalized gravitational potentials (the latter, in principle, being
produced both by the central object and the accretion disc). Finally, both
the equilibrium \emph{effective electric field }$\mathbf{E}_{s}^{eff}$\emph{,%
} generated by the combined action of the effective EM potentials and
defined as 
\begin{equation}
\mathbf{E}_{s}^{eff}\equiv -\nabla \Phi _{s}^{eff},  \label{ELECTRIC FIELD}
\end{equation}%
and the magnetic field $\mathbf{B}$ are also assumed to be axi-symmetric.

\section{First integrals of motion and guiding-center adiabatic invariants}

In the present formulation, assuming axi-symmetry and stationary EM and
gravitational fields, the exact first integrals of motion can be immediately
recovered from the symmetry properties of the single charged particle
Lagrangian function $\mathscr{L}$. In particular, these are the total
particle energy%
\begin{equation}
\left. E_{s}=\frac{M_{s}}{2}v^{2}\mathbf{+}Z_{s}e\Phi _{s}^{eff}(\mathbf{r}%
),\right.  \label{total_energy}
\end{equation}%
and the canonical momentum $p_{\varphi s}$ (conjugate to the ignorable
toroidal angle $\varphi $)%
\begin{equation}
p_{\varphi s}=M_{s}R\mathbf{v\cdot e}_{\varphi }+\frac{Z_{s}e}{c}\psi \equiv 
\frac{Z_{s}e}{c}\psi _{\ast s}.  \label{p_fi}
\end{equation}%
Gyrokinetic theory allows one to derive the adiabatic invariants of the
system \cite{Catto1977,Catto1978}; by construction, these are quantities
conserved only in an asymptotic sense, i.e., only to a prescribed order of
accuracy. As is well known, gyrokinetic theory is a basic prerequisite for
the investigation both of kinetic instabilities (see for example \cite%
{Berk1967,Tang1992,Bondeson1994}) and of equilibrium flows occurring in
magnetized plasmas \cite%
{Catto1987,Hinton1985,Tessarotto1992,Tessarotto1993,Tessarotto1994}. For
astrophysical plasmas close to compact objects, this generally involves the
treatment of strong gravitational fields which needs to be based on a
covariant formulation (see \cite{Beklemishev1999,Beklemishev2004,
Cremaschini2006,Cremaschini2008}). However, for non-relativistic plasmas (in
the sense already discussed), the appropriate formulation can also be
directly recovered via a suitable reformulation of the standard
(non-relativistic) theory for magnetically confined laboratory plasmas \cite%
{Catto1978,Littlejohn1979,Littlejohn1981,Littlejohn1983,Dubin1983,Hahm1988,Bernstein1985,Bernstein1986,Balescu,Meiss1990}%
. In connection with this, consider again the Lagrangian function $%
\mathscr{L}$ of charged particle dynamics. By performing a gyrokinetic
transformation of $\mathscr{L}$, accurate to the prescribed order in $%
\varepsilon _{M}$, it follows that - by construction - the transformed
Lagrangian $\mathscr{L^{\prime }}$ becomes independent of the guiding-center
gyrophase angle $\phi ^{\prime }.$ Therefore, by construction, the canonical
momentum $p_{\phi ^{\prime }s}^{\prime }=\partial \mathscr{L^{\prime }}%
/\partial \overset{\cdot }{\phi ^{\prime }},$ as well as the related
magnetic moment defined as$\ m_{s}^{\prime }\equiv \frac{Z_{s}e}{M_{s}c}%
p_{\phi ^{\prime }s}^{\prime },$\ are adiabatic invariants. As shown by
Kruskal (1962 \cite{Kruskal}) it is always possible to determine $%
\mathscr{L^{\prime }}$ so that $m_{s}^{\prime }$ is an adiabatic invariant
of arbitrary order in $\varepsilon _{M},$ in the sense that $\frac{1}{\Omega
_{cs}^{\prime }}\frac{d}{dt}\ln m_{s}^{\prime }=0+O(\varepsilon _{M}^{n+1}),$
where $\Omega _{cs}^{\prime }=Z_{s}eB^{\prime }/M_{s}c$ denotes the Larmor
frequency evaluated at the guiding-center and the integer $n$ depends on the
approximation used in the perturbation theory to evaluate $m_{s}^{\prime }$.
In addition, the guiding-center invariants corresponding to $E_{s}$ and $%
\psi _{\ast s}$ (denoted as $E_{s}^{\prime }$ and $\psi _{\ast s}^{\prime }$
respectively) can also be given in terms of $\mathscr{L^{\prime }}$. These
are also, by definition, manifestly independent of $\phi ^{\prime }$.,

This basic property of the magnetic moment $m_{s}^{\prime }$ is essential in
the subsequent developments. Indeed, we shall prove that it allows the
effects of temperature anisotropy to be included in the asymptotic
stationary solution.

Let us now define the concept of \emph{gyrokinetic} and \emph{equilibrium}
KDFs.

\textbf{Def. - Gyrokinetic KDF (GK KDF)}

\textit{A\ generic KDF} $f_{s}\left( \mathbf{r},\mathbf{v},t\right) $ 
\textit{will be referred to as gyrokinetic if its Lagrangian time-derivative}
$\frac{d}{dt}f_{s}\left( \mathbf{r},\mathbf{v},t\right) $ \textit{is
independent of the gyrophase angle }$\phi ^{\prime }$\textit{\ evaluated the
guiding-center position when its state }$\mathbf{x=}\left( \mathbf{r},%
\mathbf{v}\right) $ \textit{is espressed as a function of an arbitrary
gyrokinetic state }$\mathbf{z}^{\prime }\mathbf{=}\left( \mathbf{y}^{\prime
},\phi ^{\prime }\right) .$ \textit{More generally, in the following }$%
f_{s}\left( \mathbf{r},\mathbf{v},t\right) $ \textit{will be referred to as
an asymptotic-GK KDF if, neglecting corrections of} \textit{order} $O\left(
\varepsilon _{M}^{n+1}\right) ,$ $\frac{d}{dt}f_{s}\left( \mathbf{r},\mathbf{%
v},t\right) $ \textit{is independent of} $\phi ^{\prime }$\textit{.}

\textbf{Def. - Equilibrium KDF }

\textit{A\ generic KDF} $f_{s}\left( \mathbf{r},\mathbf{v},t\right) $ 
\textit{will be referred to as an equilibrium KDF if it identically
satisfies the Vlavov equation }$\frac{d}{dt}f_{s}\left( \mathbf{r},\mathbf{v}%
,t\right) =0$ and if $f_{s}$ \textit{is also independent of time, namely }$%
f_{s}=f_{s}\left( \mathbf{r},\mathbf{v}\right) .$ \textit{More generally, }$%
f_{s}\left( \mathbf{r},\mathbf{v},t\right) $ \textit{will be referred to as
an asymptotic-equilibrium KDF if, neglecting corrections of} \textit{order} $%
O\left( \varepsilon _{M}^{n+1}\right) ,$ $\frac{d}{dt}f_{s}\left( \mathbf{r},%
\mathbf{v},t\right) =0$ \textit{and to the same order} $f_{s}$ \textit{is
independent of} $t$\textit{.}

Let us first provide an example of a \emph{GK equilibrium KDF}. This can be
obtained by assuming that $f_{s}$ depends only on the exact invariants,
namely that it is of the form $f_{s}\equiv f_{\ast s}\left( E_{s},\psi
_{\ast s}\right) ,$ with $f_{\ast s}$ suitably prescribed and strictly
positive. On the other hand, an \emph{asymptotic GK equilibrium KDF }is
manifestly of the form $f_{s}\equiv \widehat{f_{\ast s}}\left( E_{s},\psi
_{\ast s},m_{s}^{\prime }\right) $ [again to be assumed as strictly positive]%
$.$ In fact, in this case, by construction, the KDF is an adiabatic
invariant of prescribed order $n$, such that 
\begin{equation}
\frac{1}{\Omega _{cs}^{\prime }}\frac{d}{dt}\ln \widehat{f_{\ast s}}%
=0+O\left( \varepsilon _{M}^{n+1}\right) ,  \label{A VLASOV}
\end{equation}%
(\textit{asymptoptic Vlasov equation)}. In particular, the order $n$ (with $%
n\geq 0$) can in principle be selected at will. We stress, however, that
since gyrokinetic theory is intrinsically asymptotic any GK equilibrium KDF
depending on the magnetic moment $m_{s}^{\prime }$ is necessarily asymptotic
in the sense of the previous two definitions.

Regarding the notations used in the following, we remark that, unless
differently specified: 1) the symbol \textquotedblleft $\wedge $%
\textquotedblright\ denotes physical quantities which refer to the treatment
of anisotropic temperatures; 2) the symbol \textquotedblleft $\ast $%
\textquotedblright\ is used to denote variables which depend on the
canonical momentum $\psi _{\ast s}$.

\section{Asymptotic equilibria with non-isotropic temperature}

In this section we derive an equilibrium solution for the KDF describing AD
plasmas for which the temperature is anisotropic.

Let us first assume that the AD plasma is characterized by a mainly toroidal
flow velocity, where the toroidal component is expressed in terms of the
angular frequency by $\mathbf{V}_{s}\left( R,z\right) \cdot \mathbf{e}%
_{\varphi }\equiv R\Omega _{s}\left(R,z\right) $. In the following, we will
also require that the plasma is locally characterized by a family of nested
magnetic surfaces which close inside the plasma in such a way that any $\psi
=const.$ surface is a closed one. We also assume that the system can be
satisfactorily described by a closed set of fluid equations in terms of four
moments of the KDF, giving the number density, the flow velocity and the
parallel and perpendicular temperatures.

The presence of a temperature anisotropy means that the plasma KDF cannot be
a Maxwellian. As already mentioned, it remains in principle completely
unspecified, with just the constraint that it must be a function only of the
first integrals of motion or the adiabatic invariants of the system. Any
non-negative KDF depending on the constants of motion and the adiabatic
invariants is therefore an acceptable solution. This freedom in choosing a
stationary solution is a well-known property of the Vlasov equation. We will
show that, in these circumstances, it is still possible to construct a
satisfactory asymptotic GK equilibrium KDF (in the sense defined in Sec.
III) which is an adiabatic invariant expressed in terms of the two first
integrals of motion (\ref{total_energy}), (\ref{p_fi}) and the
guiding-center magnetic moment $m_{s}^{\prime }$. In particular, the form of
the stationary KDF which we are going to introduce is characterized by the
following properties: 1) it is analytically tractable; 2) it affords an
explicit determination of the relevant kinetic constraints to be imposed on
the fluid fields (see the discussion after Eq.(\ref{H*})); 3) it represents a
possible kinetic model which is consistent with fluid descriptions of
collisionless plasmas characterized by temperature anisotropy; 4) it is
suitable for comparisons with previous literature, in which astrophysical
plasmas have been treated by means of a Maxwellian or a bi-Maxwellian KDF
(see for example \cite{Quataert2002, Quataert2007B, Snyder1997}). Then,
following \cite{Catto1987, Cremasch2008-2}, a convenient solution is given
by 
\begin{eqnarray}
\widehat{f_{\ast s}} &=&\frac{\widehat{\beta _{\ast s}}}{\left( 2\pi
/M_{s}\right) ^{3/2}\left( T_{\parallel \ast s}\right) ^{1/2}}
\label{f*_twotemp} \\
&&\times \exp \left\{ -\frac{H_{\ast s}}{T_{\parallel \ast s}}-m_{s}^{\prime
}\widehat{\alpha _{\ast s}}\right\}  \notag
\end{eqnarray}%
(\emph{Generalized bi-Maxwellian KDF}), where 
\begin{eqnarray}
\widehat{\beta _{\ast s}} &\equiv &\frac{\eta _{s}}{\widehat{T}_{\perp s}},
\\
\widehat{\alpha _{\ast s}} &\equiv &\frac{B^{\prime }}{\widehat{\Delta
_{T_{s}}}},  \label{alfastar} \\
H_{\ast s} &\equiv &E_{s}-\frac{Z_{s}e}{c}\psi _{\ast s}\Omega _{\ast s},
\label{H*}
\end{eqnarray}%
while $E_{s}$ is given by Eq.(\ref{total_energy}), $\psi _{\ast s}$ is given
by Eq.(\ref{p_fi}) and $\frac{1}{\widehat{\Delta _{T_{s}}}}\equiv \frac{1}{%
\widehat{T}_{\perp s}}-\frac{1}{T_{\parallel \ast s}}$. In order for the
solution (\ref{f*_twotemp}) to be a function of the integrals of motion and
the adiabatic invariants, the functions $\widehat{\beta _{\ast s}}$, $%
\widehat{\alpha _{\ast s}}$, $T_{\parallel \ast s}$ and $\Omega _{\ast s}$
must depend on the constants of motion by themselves. In general this would
require a functional dependence on both the total particle energy and the
canonical momentum. However, in the following, for simplicity, we shall
consider the case in which only a dependence on $\psi _{\ast s}$ is retained 
\cite{Catto1987, Cremasch2008-2}. Namely, $\widehat{f_{\ast s}}$ depends, by
assumption, on the flux functions $\left\{ \widehat{\beta _{\ast s}}%
,T_{\parallel \ast s},\widehat{\alpha _{\ast s}},\Omega _{\ast s}\right\} :$%
\begin{eqnarray}
\widehat{\beta _{\ast s}} &=&\widehat{\beta _{\ast s}}\left( \psi _{\ast
s}\right) ,  \label{kinconst1} \\
T_{\parallel \ast s} &=&T_{\parallel \ast s}\left( \psi _{\ast s}\right) , \\
\widehat{\alpha _{\ast s}} &=&\widehat{\alpha _{\ast s}}\left( \psi _{\ast
s}\right) , \\
\Omega _{\ast s} &=&\Omega _{\ast s}(\psi _{\ast s}),  \label{kinconst4}
\end{eqnarray}%
which in the following will be referred to as \emph{kinetic constraints}.
From these considerations it is clear that the KDF $\widehat{f_{\ast s}}$ is
itself an adiabatic invariant, and is therefore an asymptotic solution of
the stationary Vlasov equation, whose order of accuracy is uniquely
determined by the magnetic moment, as already anticipated.

From definition (\ref{H*}), it follows immediately that an equivalent
representation for $\widehat{f_{\ast s}}$ is given by:%
\begin{eqnarray}
\widehat{f_{\ast s}} &=&\frac{\widehat{\beta _{\ast s}}\exp \left[ \frac{%
X_{\ast s}}{T_{\parallel \ast s}}\right] }{\left( 2\pi /M_{s}\right)
^{3/2}\left( T_{\parallel \ast s}\right) ^{1/2}}  \label{f*-bim} \\
&&\times \exp \left\{ -\frac{M_{s}\left( \mathbf{v}-\mathbf{V}_{\ast
s}\right) ^{2}}{2T_{\parallel \ast s}}-m_{s}^{\prime }\widehat{\alpha _{\ast
s}}\right\} ,  \notag
\end{eqnarray}%
where $\mathbf{V}_{\ast s}=\mathbf{e}_{\varphi }R\Omega _{\ast s}(\psi
_{\ast s})$ and%
\begin{equation}
X_{\ast s}\equiv \left( M_{s}\frac{\left\vert \mathbf{V}_{\ast s}\right\vert
^{2}}{2}+\frac{Z_{s}e}{c}\psi \Omega _{\ast s}-Z_{s}e\Phi _{s}^{eff}\right) .
\end{equation}

The same kinetic constraints (\ref{kinconst1})-(\ref{kinconst4}) also apply
to the solution (\ref{f*-bim}). Note that the functions $\widehat{\beta
_{\ast s}}\exp \left[ \frac{X_{\ast s}}{T_{\parallel \ast s}}\right] ,$ $%
\mathbf{V}_{\ast s}$ and $T_{\parallel \ast s}$ cannot be regarded as \emph{%
fluid fields}, since they have a dependence on the particle velocity via the
canonical momentum $\psi _{\ast s}$. On the other hand, fluid fields must be
computed as integral moments of the distribution function over the particle
velocity $\mathbf{v}$.

Next we show that a convenient asymptotic expansion for the adiabatic
invariant $\widehat{f_{\ast s}}$ can be properly obtained in the following
suitable limit. Consider, in fact, the quantity $\varepsilon $ defined as $%
\varepsilon \equiv \max \left\{ \varepsilon _{s},s=i,e\right\} $, with $%
\varepsilon _{s}\equiv \left\vert \frac{L_{\varphi s}}{p_{\varphi
s}-L_{\varphi s}}\right\vert =\left\vert \frac{M_{s}Rv_{\varphi }}{\frac{%
Z_{s}e}{c}\psi }\right\vert $, where we have used the definition (\ref{p_fi}%
) with $v_{\varphi }\equiv \mathbf{v\cdot e}_{\varphi }$, and where $%
L_{\varphi s}$ denotes the species particle angular momentum. We can give an
average upper limit estimate for the magnitude of $\varepsilon _{s}$ in
terms of the species thermal velocity and the inverse aspect ratio
previously defined. To do this, we first set $\psi \sim B_{P}r^{2}$, which
is appropriate for the domain of closed nested magnetic surfaces. Recall
that here $r$ is the average poloidal radius of a generic nested magnetic
surface. In this evaluation, the species thermal velocities $v_{ths}$ and
the toroidal flow velocities $R\Omega _{s}$ are considered to be of the same order
with respect to the $\varepsilon $-expansion, i.e. $v_{ths}/R\Omega _{s}\sim
O\left( \varepsilon ^{0}\right) $ (referred to as \textit{sonic flow}). Therefore, assuming 
$v_{\varphi s}\sim v_{ths}$ it follows immediately that $\varepsilon
_{s}\sim \frac{r_{Ls}}{L_{C}}$, where $r_{Ls}$ is the species Larmor radius
and $L_{C}\equiv r\delta $, with $\delta $ the inverse aspect ratio. We
shall say that the AD plasma is \emph{strongly magnetized} whenever $%
0<\varepsilon \ll 1$. This condition is realized if $r\geq r_{\min }$, where 
$r_{\min }=\max \left\{ \frac{r_{Ls}}{\varepsilon _{s}\delta },s=i,e\right\} 
$ is the minimum average poloidal radius of the toroidal nested magnetic
surfaces for which $\varepsilon \ll 1$ is satisfied. In this case $%
\varepsilon $ can be taken as a small parameter for making a Taylor
expansion of the KDF and its related quantities, by setting $\psi _{\ast
s}\simeq \psi +O\left( \varepsilon ^{k}\right) $, $k\geq 1$. From the above
discussion, it is clear that this asymptotic expansion is valid for $r$
within an interval $r_{\min }\leq r\leq r_{\max }$, where the lower bound is
fixed by the condition of having a strongly magnetized plasma, while the
upper bound is given by the geometric properties of the system and the small
inverse aspect ratio ordering. For the purpose of this paper, in performing
the asymptotic expansion we retain the leading-order expression for the
guiding-center magnetic moment $m_{s}^{\prime }\simeq \mu _{s}^{\prime }=%
\frac{M_{s}w^{\prime 2}}{2B^{\prime }}$ \cite{Kruskal}. Then, it is
straightforward to prove that for strongly magnetized plasmas, the following
relation holds to first order in $\varepsilon $ (i.e., retaining only linear
terms in the expansion): $\widehat{f_{\ast s}}=\widehat{f_{s}}\left[ 1+h_{Ds}%
\right] +O\left( \varepsilon ^{n}\right) $, $n\geq 2$. Here, the zero order
distribution $\widehat{f_{s}}$ is expressed as 
\begin{eqnarray}
\widehat{f_{s}} &=&\frac{n_{s}}{\left( 2\pi /M_{s}\right) ^{3/2}\left(
T_{\parallel s}\right) ^{1/2}T_{\perp s}}  \label{bimaxz} \\
&&\times \exp \left\{ -\frac{M_{s}\left( \mathbf{v}-\mathbf{V}_{s}\right)
^{2}}{2T_{\parallel s}}-\frac{M_{s}w^{\prime 2}}{2\Delta _{T_{s}}}\right\} ,
\notag
\end{eqnarray}%
which we will here call the \emph{bi-Maxwellian KDF}, where $\frac{1}{\Delta
_{T_{s}}}\equiv \frac{1}{T_{\perp s}}-\frac{1}{T_{\parallel s}}$, the number
density $n_{s}=\eta _{s}\exp \left[ \frac{X_{s}}{T_{\parallel s}}\right] $
and%
\begin{equation}
X_{s}\equiv \left( M_{s}\frac{R^{2}\Omega _{s}^{2}}{2}+\frac{Z_{s}e}{c}\psi
\Omega _{s}(\psi )-Z_{s}e\Phi _{s}^{eff}\right) ,  \label{X}
\end{equation}%
with $\eta _{s}$ denoting the \emph{pseudo-density}. Then, $\mathbf{V}_{s}=%
\mathbf{e}_{\varphi }R\Omega _{s}$ and the following kinetic\emph{\ }%
constraints are implied from (\ref{kinconst1})-(\ref{kinconst4}): $\beta
_{s}=\beta _{s}\left( \psi \right) =\frac{\eta _{s}}{T_{\perp s}}$, $%
T_{\parallel s}=T_{\parallel s}\left( \psi \right) $, $\widehat{\alpha _{s}}=%
\widehat{\alpha _{s}}\left( \psi \right) =\frac{B^{\prime }}{\Delta _{T_{s}}}
$, $\Omega _{s}=\Omega _{s}(\psi )$. As can be seen, the functional form of
the leading order number density, the flow velocity and the temperatures
carried by the bi-Maxwellian KDF is naturally determined. In particular,
note that the flow velocity is species-dependent, while the related angular
frequency $\Omega _{s}$ must necessarily be constant on each nested toroidal
magnetic surface $\left\{ \psi (\mathbf{r})=const.\right\} $. Finally, the
quantity $h_{Ds}$ represents the \emph{diamagnetic part} of the KDF $%
\widehat{f_{\ast s}}$, given by%
\begin{equation}
h_{Ds}=\left\{ \frac{cM_{s}R}{Z_{s}e}Y_{1}+\frac{M_{s}R}{T_{\parallel s}}%
Y_{2}\right\} \left( \mathbf{v\cdot }\mathbf{e}_{\varphi }\right) ,
\label{hdd}
\end{equation}%
with $Y_{1}\equiv \left[ A_{1s}+A_{2s}\left( \frac{H_{s}}{T_{\parallel s}}-%
\frac{1}{2}\right) -\mu _{s}^{\prime }\widehat{A_{4s}}\right] $, $%
H_{s}\equiv E_{s}-\frac{Z_{s}e}{c}\psi _{s}\Omega _{s}(\psi _{s})$ and $%
Y_{2}\equiv \Omega _{s}(\psi )\left[ 1+\psi A_{3s}\right] $, where we have
introduced the following definitions: $A_{1s}\equiv \frac{\partial \ln \beta
_{s}}{\partial \psi },$ $A_{2s}\equiv \frac{\partial \ln T_{\parallel s}}{%
\partial \psi },$ $A_{3s}\equiv \frac{\partial \ln \Omega _{s}(\psi )}{%
\partial \psi },$ $\widehat{A_{4s}}\equiv \frac{\partial \widehat{\alpha _{s}%
}}{\partial \psi }.$ We remark here that: 1) in the $\varepsilon $-expansion
of (\ref{f*_twotemp}), performed around (\ref{bimaxz}), no magnetic or
electric field scale lengths enter, as can be seen from Eq.(\ref{hdd}); 2)
we also implicitly assume the validity of the ordering $\frac{\varepsilon
_{M}}{\varepsilon }\ll 1$, which will be discussed below (see next section). For
this reason, corrections of $O\left( \varepsilon _{M}^{k}\right) $, with $%
k\geq 1$, to (\ref{hdd}) have been neglected; 3) in this $%
\varepsilon -$expansion we have also assumed that the scale-length $L$ is of
the same order (with respect to $\varepsilon $) as the characteristic scale-lengths associated with the
species pseudo-densities $\eta _{s}$, the temperatures $T_{\parallel s}$ and $%
T_{\perp s}$, and the toroidal rotational frequencies $\Omega _{s}$.

To conclude this section we point out that the very existence of the present
asymptotic kinetic equilibrium solution and the realizability of the kinetic
constraints implied by it, must be checked for consistency also with the
constraints imposed by the Maxwell equations, as discussed in Sec. VII.

\section{Moments of the KDF}

It is well known that, given a distribution function, it is always possible
to compute the fluid moments associated with it, which are defined through
integrals of the distribution over the velocity space. Although an exact
calculation of the fluid moments could be carried out (e.g., numerically)
for prescribed kinetic closures, in this section we want to take advantage
of the asymptotic expansion of the KDF in the limit of strongly magnetized
plasmas to evaluate them analytically, thanks to the properties of the
bi-Maxwellian KDF. In the following, we provide approximate expressions for
the number density and the flow velocity, which allow one to write the
Poisson and Ampere equations for the EM fields in a closed form, and for the
non-isotropic species pressure tensor. Since these fluid fields are then
known (in terms of suitable kinetic flux functions and with a prescribed
accuracy), the closure problem characteristic of the fluid theories is then
naturally solved as well.

The main feature of this calculation is that the number density and flow
velocity are computed by performing a transformation of all of the
guiding-center quantities appearing in the asymptotic equilibrium KDF to the
actual particle position, to leading order in $\varepsilon _{M}$ (according
to the order of accuracy of the adiabatic invariant used), and they are then
determined up to first order in $\varepsilon $, in agreement with the order
of expansion previously set for the KDF. Terms of higher order, i.e. $%
O\left( \varepsilon _{M}^{n}\right) ,$ with $n\geq 1$, and $O\left(
\varepsilon ^{k}\right) ,$ with $k\geq 2$, as well as mixed terms of order $%
O\left( \left[ \varepsilon \varepsilon _{M}\right] ^{n}\right) $ with $n\geq
1$, are therefore neglected in the present calculation. This approximation
clearly holds if $\frac{\varepsilon _{M}}{\varepsilon }\ll 1$, which is
consistent with the present assumptions. In fact, from the definitions given
for these two small dimensionless parameters it follows that%
\begin{equation}
\frac{\varepsilon _{M}}{\varepsilon }\sim O\left( \delta \right) \ll 1.
\end{equation}

To first order in $\varepsilon $, the total number density $n_{s}^{tot}$ is
given by $n_{s}^{tot}\equiv \int d\mathbf{v}\widehat{f_{\ast s}}\simeq n_{s}%
\left[ 1+\Delta _{n_{s}}\right] $. Note here that the resulting number
density has two distinct contributions: $n_{s}$ is the zero order term given
in the previous section, while $\Delta _{n_{s}}$ represents the term of $%
O\left( \varepsilon \right) $ which carries all of the corrections due to
the asymptotic expansion of the KDF for strongly magnetized plasmas. The
full expression for $\Delta _{n_{s}}$ is given in Appendix A. Finally, a
similar integral can be performed to compute the total flow velocity $%
\mathbf{V}_{s}^{tot}$. This has the form $n_{s}^{tot}\mathbf{V}%
_{s}^{tot}\equiv \int d\mathbf{vv}\widehat{f_{\ast s}}\simeq n_{s}\left[ 
\mathbf{V}_{s}+\Delta \mathbf{U}_{s}\right] ,$ where by definition $\mathbf{V%
}_{s}=\Omega _{s}(\psi )R\mathbf{e}_{\varphi }$ and $\Delta \mathbf{U}_{s}$
represents the \emph{self-consistent FLR velocity corrections} given by:%
\begin{equation}
\Delta \mathbf{U}_{s}\equiv \Delta _{\varphi s}\mathbf{e}_{\varphi }+\frac{%
\Delta _{3s}}{B}\nabla \psi \times \nabla \varphi ,  \label{delta-flow}
\end{equation}%
where $\Delta _{\varphi s}\equiv \Delta _{n_{s}}\Omega _{s}R+\Delta
_{2s}+\Delta _{3s}\frac{I}{RB}$. Note that in Eq.(\ref{delta-flow}) the
terms proportional to $\Delta _{n_{s}}$, $\Delta _{2s}$ and $\Delta _{3s}$
come from the asymptotic expansion of the KDF for strongly magnetized
plasmas and are of $O\left( \varepsilon \right) $ with respect to the
toroidal velocity $\Omega _{s}R$. The full expressions for $\Delta _{n_{s}}$%
, $\Delta _{2s}$ and $\Delta _{3s}$ are given in Appendix A. The first-order
term $\Delta \mathbf{U}_{s}$ provides corrections to the zero-order toroidal
flow velocity with components in all of the three space directions and so we
can conclude that, although the dominant fluid velocity is mainly toroidal,
there is also a poloidal component of order $\varepsilon $, associated with
the term $\frac{\Delta _{3s}}{B}\nabla \psi \times \nabla \varphi $.
However, this is not necessarily an accretion velocity, especially under the
hypothesis of closed nested magnetic surfaces which define a local domain in
which the disc plasma is confined. Moreover, note that the ratio between the
toroidal and poloidal velocities depends also on $\delta _{Ts}$, in the
sense that $\frac{\left\vert \frac{\Delta _{3s}}{B}\nabla \psi \times \nabla
\varphi \right\vert }{\left\vert \mathbf{V}_{s}\right\vert }\sim O\left(
\varepsilon \right) O\left( \delta _{Ts}\right) $. The magnitude of the
temperature anisotropy can therefore be relevant in further decreasing the
poloidal velocity in comparison with the toroidal one, which on the contrary
is not affected by $\delta _{Ts}$. However, the real importance of this
result in connection with the astrophysics of collisionless AD plasmas is,
instead, the fact that this poloidal velocity is a primary source for a
poloidal current density which in turn can generate a finite toroidal
magnetic field (see the section on the Maxwell equations). This means that,
even without any net accretion of disc material (which would require at
least a redistribution of the angular momentum), the kinetic equilibrium
solution provides a mechanism for the generation of a toroidal magnetic
field, with serious implications for the stability analysis of these
equilibria. The physical mechanism responsible for this poloidal drift is
purely kinetic and is essentially due to the conservation of the canonical
toroidal momentum and the FLR effects associated with the temperature
anisotropy. As a last point, consider the species non-isotropic pressure
tensor, which is defined by the following moment of the KDF: $\underline{%
\underline{\Pi }}_{s}=\int d\mathbf{v}M_{s}(\mathbf{v}-\mathbf{V}_{s}^{tot})(%
\mathbf{v}-\mathbf{V}_{s}^{tot})\widehat{f_{\ast s}}.$ Then, the overall
pressure tensor of the system is obtained by summing the single species
pressure tensors: $\underline{\underline{\Pi }}=\sum_{s=i,e}\underline{%
\underline{\Pi }}_{s}$. A direct analytical calculation retaining only the
zero-order terms (with respect to all of the small dimensionless parameters)
in the Taylor expansion of the KDF $\widehat{f_{\ast s}}$ (whose
leading-order expression coincides with the bi-Maxwellian KDF), shows that 
\textit{the corresponding species tensor pressure is non-isotropic} to
leading order and in this approximation is given by:%
\begin{equation}
\underline{\underline{\Pi }}_{s}=p_{\perp s}\underline{\mathbf{I}}+\left(
p_{\parallel s}-p_{\perp s}\right) \mathbf{bb},  \label{tensor-pressure anis}
\end{equation}%
where $p_{\perp s}\equiv n_{s}T_{\perp s}$ and $p_{\parallel s}\equiv
n_{s}T_{\parallel s}$ represent the leading-order perpendicular and parallel
pressures. The divergence of the species pressure tensor is of particular
interest; this is given by:%
\begin{equation}
\nabla \cdot \underline{\underline{\Pi }}_{s}=\nabla p_{\perp s}+\mathbf{bB}%
\cdot \nabla \left( \frac{p_{\parallel s}-p_{\perp s}}{B}\right) -\Delta
p_{s}\mathbf{Q},
\end{equation}%
where $\mathbf{Q\equiv }\left[ \mathbf{bb\cdot }\nabla \ln B+\frac{4\pi }{cB}%
\mathbf{b\times J}-\nabla \ln B\right] $ and $\Delta p_{s}\equiv \left(
p_{\parallel s}-p_{\perp s}\right) $.

\section{The case of isotropic temperature}

In this section, we consider the case of isotropic temperature for the
equilibrium distribution $\widehat{f_{\ast s}}$. When the condition $%
T_{\parallel s}=T_{\perp s}\equiv T_{s}$ is satisfied, the stationary KDF
reduces to $f_{\ast s},$ where%
\begin{equation}
f_{\ast s}=\frac{\eta _{\ast s}}{\pi ^{3/2}\left( 2T_{\ast s}/M_{s}\right)
^{3/2}}\exp \left\{ -\frac{H_{\ast s}}{T_{\ast s}}\right\}  \label{maxgen}
\end{equation}%
is referred to as the \textit{Generalized Maxwellian Distribution with
isotropic temperature }\cite{Cremasch2008-2}. Here, $H_{\ast s}$ retains its
definition (\ref{H*}), while the kinetic constraints are expressed for the
quantities $\widehat{n}_{\ast s}$ and $T_{\ast s}$, whose functional
dependence is $\eta _{\ast s}=\eta _{s}(\psi _{\ast s})$ and $T_{\ast
s}=T_{s}(\psi _{\ast s})$. By construction, this distribution function is
expressed only in terms of the first integrals of motion of the system and
is therefore an \emph{exact kinetic equilibrium} solution. Performing an
asymptotic expansion in the limit of strong magnetic field, as done before
for $\widehat{f_{\ast s}}$, gives the following result: $f_{\ast s}=f_{Ms}%
\left[ 1+h_{Ds}\right] +O(\varepsilon ^{n}),$ $n\geq 2$, where%
\begin{equation}
f_{Ms}=\frac{n_{s}}{\pi ^{3/2}\left( 2T_{s}/M_{s}\right) ^{3/2}}\exp \left\{
-\frac{M_{s}\left( \mathbf{v}-\mathbf{V}_{s}\right) ^{2}}{2T_{s}}\right\}
\end{equation}%
is the zero-order term of the series, which coincides with a drifted
Maxwellian KDF with $T_{s}=T_{s}\left( \psi \right) $, $\mathbf{V}%
_{s}=\Omega _{s}(\psi )R\mathbf{e}_{\varphi }$ and $n_{s}=\eta _{s}\left(
\psi \right) \exp \left[ \frac{X_{s}^{\sim }}{T_{s}}\right] $. In this case,
the function $h_{Ds}$ is given by%
\begin{equation}
h_{Ds}=\left\{ \frac{cM_{s}R}{Z_{s}e}Y_{1}+\frac{M_{s}R}{T_{s}}Y_{2}\right\}
\left( \mathbf{v\cdot }\mathbf{e}_{\varphi }\right) ,
\end{equation}%
with $Y_{1}\equiv \left[ A_{1s}+A_{2s}\left( \frac{H_{s}}{T_{s}}-\frac{3}{2}%
\right) \right] $ and $Y_{2}\equiv \Omega _{s}(\psi )\left[ 1+\psi A_{3s}%
\right] $, where $A_{1s}\equiv \frac{\partial \ln \eta _{s}}{\partial \psi }%
, $ $A_{2s}\equiv \frac{\partial \ln T_{s}}{\partial \psi },$ $A_{3s}\equiv 
\frac{\partial \ln \Omega _{s}(\psi )}{\partial \psi }$. Finally, as shown
in \cite{Cremasch2008-2}, the angular frequency is given to leading order by 
$\Omega _{s}(\psi )=\frac{\partial \left\langle \chi \right\rangle }{%
\partial \psi },$ where $\chi \equiv c\Phi _{s}^{eff}+\frac{cT_{s}}{Z_{s}e}%
\ln n_{s}$.

Before concluding this section, we stress again that the case with isotropic
temperature represents a result whose accuracy is not limited by dependence
on any gyrokinetic invariant and that it does not require any guiding-center
variable transformation. For this reason, there are no restrictions of
applicability of the solution (\ref{maxgen}) which, in principle, holds also
in the limit of vanishing magnetic field.

\section{The Maxwell equations and the \textquotedblleft kinetic
dynamo\textquotedblright}

In this section we write the Poisson and Ampere equations for the EM fields
explicitly, pointing out the consequences of the kinetic treatment developed
in Sections IV-VI. In particular we prove that, besides a self-generated
poloidal magnetic field, the kinetic equilibrium can sustain also a toroidal
field (which may be thought of as being a \emph{kinetic dynamo}), thanks to
the combined effects of FLR corrections and temperature anisotropies. For
definiteness, let us consider the Poisson equation for the electrostatic
potential $\Phi ,$ expressed as%
\begin{equation}
\nabla ^{2}\Phi =-4\pi \sum_{s=i,e}q_{s}n_{s}\left[ 1+\Delta _{n_{s}}\right]
,  \label{Poisson-1}
\end{equation}%
where $\Delta _{n_{s}}$ is written out explicitly in Appendix A. In the
limit of a strongly magnetized plasma and considering the accuracy of the
previous asymptotic analytical expansions, we shall say that the plasma is 
\textit{quasi-neutral} if the ordering $\frac{-\nabla ^{2}\Phi }{4\pi
\sum_{s=i,e}q_{s}n_{s}\left[ 1+\Delta _{n_{s}}\right] }=0+O\left(
\varepsilon ^{k}\right) $, with $k\geq 2$ holds, whereas we call it \textit{%
weakly non-neutral} if $\frac{-\nabla ^{2}\Phi }{4\pi \sum_{s=i,e}q_{s}n_{s}%
\left[ 1+\Delta _{n_{s}}\right] }=0+O\left( \varepsilon \right) .$ The
kinetic equilibrium for a weakly non-neutral plasma is referred to as a 
\textit{Hall kinetic equilibrium }\cite{Cremasch2008-2}, and the
corresponding fluid configuration is referred to as a \textit{Hall
Gravitational MHD (Hall-GMHD) fluid equilibrium }\cite{Cremasch2008-1}.

Next, we show that quasi-neutrality (in the sense just defined) can be
locally satisfied by imposing a suitable constraint on the electrostatic
(ES) potential $\Phi $. It can be shown that this constraint can always be
satisfied since the leading-order contribution to the ES potential remains
unaffected. The result follows by neglecting higher-order corrections to the
number density ($\Delta _{n_{s}}$) and setting $q_{i}=Ze$ and $q_{e}=-e$.
Thanks to the arbitrariness in the choice of the flux functions introduced
by the kinetic constraints (see Sec. IV), it is possible to show that
quasi-neutrality implies the following constraint for the oscillatory part $%
\Phi ^{\sim }$ of the ES potential, i.e., correct to both $O\left(
\varepsilon ^{0}\right) $ and $O\left( \varepsilon _{M}^{0}\right) $,%
\begin{equation}
\Phi ^{\sim }\left( \psi ,\vartheta \right) \equiv \Phi -\left\langle \Phi
\right\rangle \simeq \frac{S^{\sim }}{e\left( \frac{Z}{T_{\parallel i}}+%
\frac{1}{T_{\parallel e}}\right) },  \label{Fi}
\end{equation}%
where $S^{\sim }\equiv \ln \left( \frac{\eta _{e}}{Z\eta _{i}}\right) +\left[
\frac{\overline{X}_{e}}{T_{\parallel e}}-\frac{\overline{X}_{i}}{%
T_{\parallel i}}\right] $, and $\overline{X}_{s}\equiv \left( M_{s}\frac{%
R^{2}\Omega _{s}^{2}}{2}+\frac{Z_{s}e}{c}\psi \Omega _{s}(\psi )-M_{s}\Phi
_{G}\right) $. In particular, the arbitrariness in the coefficient $\frac{%
\eta _{e}}{Z\eta _{i}}$ can be used to satisfy the constraint $\left\langle
S^{\sim }\right\rangle =0$. In fact, in view of Eq.(\ref{kinconst1}), it
follows that%
\begin{equation}
\eta _{s}=\frac{\beta _{s}\left( \psi \right) T_{\parallel s}\left( \psi
\right) }{1+\frac{\alpha _{s}\left( \psi \right) T_{\parallel s}\left( \psi
\right) }{B\left( \psi ,\vartheta \right) }},
\end{equation}%
where $\alpha _{s}\left( \psi \right) $ is related to $\widehat{\alpha _{s}}%
\left( \psi \right) $ as outlined in Appendix A and the flux functions still
remain arbitrary. In conclusion, Eq.(\ref{Fi}) determines only $\Phi ^{\sim
} $ and not the total ES potential. Note that this solution for the
electrostatic potential $\Phi ^{\sim }$ can be shown to be consistent with
earlier treatments appropriate for Tokamak plasma equilibria \cite%
{Catto1987,Tessarotto1992}. This can be exactly recovered thanks to the
arbitrariness in defining the pseudo-densities and by taking the limit of
isotropic temperatures and zero gravitational potential, as in the case of
laboratory plasmas. In this limit the species pseudo-densities become flux
functions \cite{Catto1987}. Then, because of this arbitrariness, by taking%
\begin{equation}
\frac{\eta _{e}}{Z\eta _{i}}=1,
\end{equation}%
it follows that Eq.(\ref{Fi}) reduces to the form presented in \cite%
{Catto1987}, which can only be used to determine the poloidal variation of
the potential.

Let us now consider the Ampere equation. Adopting the Taylor analytical
expansion of the asymptotic equilibrium KDF and neglecting corrections of $%
O\left( \varepsilon ^{k}\right) $, with $k\geq 2$, and $O\left( \varepsilon
_{M}^{n+1}\right) $, with $n\geq 0$, this can be approximately written as
follows:%
\begin{equation}
\nabla \times \mathbf{B}^{self}=\frac{4\pi }{c}\sum_{s=i,e}q_{s}n_{s}\left[ 
\mathbf{V}_{s}+\Delta \mathbf{U}_{s}\right] ,  \label{Ampere-2}
\end{equation}%
where $\mathbf{B}^{self}$ is as defined in Eq.(\ref{B FIELD}) and the
expression for $\Delta \mathbf{U}_{s}$ is given by Eq.(\ref{delta-flow}).
The toroidal component of this equation gives the generalized Grad-Shafranov
equation for the poloidal flux function $\psi _{p}$:%
\begin{equation}
\Delta ^{\ast }\psi _{p}=-\frac{4\pi }{c}R\sum\limits_{s=e,i}q_{s}n_{s}\left[
\Omega _{s}\left( \psi \right) R+\Delta _{\varphi s}\right] ,  \label{GSapp}
\end{equation}%
where the elliptic operator $\Delta ^{\ast }$ is defined as $\Delta ^{\ast
}\equiv R^{2}\nabla \cdot \left( R^{-2}\nabla \right) $ \cite{meiss}. The
remaining terms in Eq.(\ref{Ampere-2}) give the equation for the toroidal
component of the magnetic field $\frac{I(\psi ,\vartheta )}{R}$. In the same
approximation, this is:%
\begin{equation}
\nabla I(\psi ,\vartheta )\times \nabla \varphi =\frac{4\pi }{c}%
\sum_{s=i,e}q_{s}n_{s}\frac{\Delta _{3s}}{B}\nabla \psi \times \nabla
\varphi ,  \label{I}
\end{equation}%
where $\Delta _{3s}$, given in Appendix A, contains the contributions of the
species temperature anisotropies. For consistency with the approximation
introduced, in the small inverse aspect ratio ordering, it follows that $%
\frac{\partial I(\psi ,\vartheta )}{\partial \vartheta }=0+O\left( \delta
^{k}\right) ,$ i.e., to leading order in $\delta $: $I=I\left( \psi \right)
+O\left( \delta ^{k}\right) $, with $k\geq 1$. This in turn also requires
that the corresponding current density in Eq.(\ref{I}) is necessarily a flux
function. Then, correct to $O(\varepsilon )$, $O(\varepsilon _{M}^{0})$ and $%
O\left( \delta ^{0}\right) ,$ the differential equation for $I\left( \psi
\right) $ becomes:%
\begin{equation}
\frac{\partial I(\psi )}{\partial \psi }=\frac{4\pi }{c}\sum%
\limits_{s=e,i}q_{s}n_{s}\frac{\Delta _{3s}}{B},  \label{CONSTARINT EQUATION}
\end{equation}%
which uniquely determines an approximate solution for the toroidal magnetic
field. This result is remarkable because it shows that there is a \textit{%
stationary \textquotedblleft kinetic dynamo\ effect\textquotedblright\ }%
which generates an equilibrium toroidal magnetic field \textit{without
requiring any net accretion} and \textit{in the absence of any possible
instability/turbulence phenomena}. This new mechanism results from poloidal
currents arising due to the FLR effects and temperature anisotropies which
are characteristic of the equilibrium KDF. We remark that the
self-generation of the stationary magnetic field is purely diamagnetic. In
particular, the toroidal component is associated with the drifts of the plasma
away from the flux surfaces. In the present formulation, possible dissipative
phenomena leading to a non-stationary self field have been ignored. Such dissipative phenomena
probably do arise in practice and could occur both in the local domain where
the equilibrium magnetic surfaces are closed and nested and elsewhere.
Temperature anisotropies are therefore an important physical property of
collisionless AD plasmas, giving a possible mechanism for producing a
stationary toroidal magnetic field. We stress that this effect disappears
altogether in the case of isotropic temperatures, as demonstrated in
Appendix A. Finally, we consider the ratio between the toroidal and poloidal
current densities ($\mathbf{J}_{T}$ and $\mathbf{J}_{P}$). In the small
inverse aspect ratio ordering, neglecting corrections of order $O\left(
\delta ^{k}\right) $ with $k\geq 1$, this provides an estimate of the
magnitude of the corresponding components of the magnetic field. In fact, in
this limit we can write $\frac{\left\vert \nabla \times \mathbf{B}%
_{T}\right\vert }{\left\vert \nabla \times \mathbf{B}_{P}\right\vert }\sim 
\frac{\left\vert \mathbf{B}_{T}\right\vert }{\left\vert \mathbf{B}%
_{P}\right\vert }\sim \frac{\left\vert \mathbf{J}_{P}\right\vert }{%
\left\vert \mathbf{J}_{T}\right\vert }$ and so conclude that, although for
the single species velocity, the ordering $\frac{\left\vert \frac{\Delta
_{3s}}{B}\nabla \psi \times \nabla \varphi \right\vert }{\left\vert \mathbf{V%
}_{s}\right\vert }\sim O\left( \varepsilon \right) O\left( \delta
_{Ts}\right) $ holds (see Sec. VI), this might no longer be the case for the
magnetic field, which instead depends on the ratio between the total
toroidal and poloidal current densities. In particular, the possibility of
having finite stationary toroidal magnetic fields is, in principle, allowed
by the present analysis, depending on the properties of the overall solution
describing the system.

\bigskip

\section{Conclusions}

Getting a complete understanding of the dynamical properties of
astrophysical accretion discs still represents a challenging task and there
are many open problems remaining to be solved before one can get a full and
consistent theoretical formulation for the physical processes involved.

The present investigation provides some important new results for
understanding the equilibrium properties of accretion discs, obtained within
the framework of a kinetic approach based on the Vlasov-Maxwell description.
The derivation presented applies for collisionless non-relativistic and
axi-symmetric AD plasmas under the influence of both gravitational and EM
fields. A wide range of astrophysical scenarios can be investigated with the
present theory, thanks to the possibility of properly setting the different
parameters which characterize the physical and geometrical properties of the
model. A possible astrophysical context is provided, for example, by
radiatively inefficient accretion flows onto black holes, where the
accreting material is thought to consist of a plasma of collisionless ions
and electrons with different temperatures, in which the dominant magnetic
field is generated by the plasma current density. We have considered here
the specific case in which the structure of the magnetic field is locally
characterized by a family of closed nested magnetic surfaces within which
the plasma has mainly toroidal flow velocity. For this, we have proved that
a kinetic equilibrium exists and can be described by a stationary KDF
expressed in terms of the exact integrals of motion and the magnetic moment
prescribed by the gyrokinetic theory, which is an adiabatic invariant. Many
interesting new results have been pointed out; the most relevant ones for
astrophysical applications are the following: 1) the possibility of
including the effects of a non-isotropic temperature in the stationary KDF;
2) the proof that the Maxwellian and bi-Maxwellian KDFs are asymptotic
stationary solutions, i.e. they can be regarded as approximate equilibrium
solutions in the limit of strongly magnetized plasmas; 3) the possibility of
computing the stationary fluid moments to the desired order of accuracy in
terms of suitably prescribed flux functions; 4) the proof that a toroidal
magnetic field can be generated in a stationary configuration even in the
absence of any net accretion flow if and only if the plasma has a
temperature anisotropy.

This last point, in particular, is of great interest because it gives a
mechanism for generating a stationary toroidal field in the disc,
independent of instabilities related to the accretion flow. The consistent
kinetic formulation developed here permits the self-generation of such a
field by the plasma itself, associated with localized poloidal
drift-currents on the nested magnetic surfaces as a consequence of
temperature anisotropies. This stationary poloidal motion is made possible
in the framework of kinetic theory by the conservation of the canonical
momentum as a result of FLR effects.

These results reveal and confirm the power of the kinetic treatment and the
necessity for adopting such a formalism in order to correctly understand the
physical phenomena occurring in accretion discs. This study may represent a
significant step forward for understanding the physical properties of
accretion discs in their kinetic equilibrium configurations, and it
motivates making further investigations of the subject aimed at extending
the present range of validity to more general physical configurations. The
conclusions presented here may also have important consequences for other
applications, and can provide a reference starting point for future work on
kinetic stability analysis of accretion discs.

\begin{acknowledgments}
Helpful discussions with Alexander Schekochihin (Oxford University, Oxford,
U.K.) are gratefully acknowledged (C.C. and J.M.). This work has been partly
developed (C.C. and M.T.) within the framework of the MIUR (Italian Ministry
for Universities and Research) PRIN Research Program \textquotedblleft
Modelli della teoria cinetica matematica nello studio dei sistemi complessi
nelle scienze applicate\textquotedblright \thinspace\ and the Trieste
Consortium for Magnetofluid Dynamics.
\end{acknowledgments}

\appendix

\section{Calculation of the fluid moments}

In this appendix we give the detailed expressions for the coefficients which
determine the fluid moments of the KDF, obtained by integrating the KDF over
the velocity space. For the number density, we have found that $%
n_{s}^{tot}=n_{s}\left[ 1+\Delta _{n_{s}}\right] $. The term $\Delta
_{n_{s}} $ is given by%
\begin{eqnarray}
\Delta _{n_{s}} &\equiv &V_{s}\left[ \gamma _{1}+\gamma _{3}\left( \frac{%
T_{\parallel s}}{M_{s}}+\frac{4T_{\perp s}}{M_{s}}+V_{s}^{2}\right) \right] +
\label{deltan} \\
&&+\frac{2\gamma _{3}I^{2}}{B^{2}}\frac{\left( T_{\parallel s}-T_{\perp
s}\right) V_{s}}{R^{2}M_{s}}-\frac{\gamma _{2}}{B}V_{s}T_{\perp s},  \notag
\end{eqnarray}%
where $V_{s}=R\Omega _{s}(\psi )$ and%
\begin{eqnarray}
\gamma _{1} &\equiv &\left\{ \frac{cM_{s}R}{Z_{s}e}K+\frac{M_{s}V_{s}}{%
T_{\parallel s}}\left[ 1+\psi A_{3s}\right] \right\} , \\
\gamma _{2} &\equiv &\left\{ \frac{cM_{s}R}{Z_{s}e}A_{4s}\right\} , \\
\gamma _{3} &\equiv &\left\{ \frac{cM_{s}^{2}R}{Z_{s}e}\frac{A_{2s}}{%
2T_{\parallel s}}\right\} ,
\end{eqnarray}%
in which $K\equiv \left[ A_{1s}+A_{2s}\left( \frac{Z_{s}e\Phi _{s}^{eff}-%
\frac{Z_{s}e}{c}\psi \Omega _{s}(\psi )}{T_{\parallel s}}-\frac{1}{2}\right) %
\right] $ and $A_{4s}\equiv \frac{\partial \alpha _{s}}{\partial \psi },$
with $\alpha _{s}\left( \psi \right) \equiv \frac{B}{\Delta _{T_{s}}}$. Note
that here $\alpha _{s}\left( \psi \right) $ differs from $\widehat{\alpha
_{s}}\left( \psi \right) $ because of the guiding-center transformation of
the magnetic field $B$.

Proceeding in the same way for evaluating the second moment of the KDF, it
can be shown that the first order correction $\Delta \mathbf{U}_{s}$ to the
toroidal flow velocity can be written as $\Delta \mathbf{U}_{s}\equiv \Delta
_{\varphi s}\mathbf{e}_{\varphi }+\frac{\Delta _{3s}}{B}\nabla \psi \times
\nabla \varphi ,$ where we recall that%
\begin{equation}
\Delta _{\varphi s}\equiv \Delta _{n_{s}}\Omega _{s}R+\Delta _{2s}+\Delta
_{3s}\frac{I}{RB}.  \label{deltaphi}
\end{equation}%
Here $\Delta _{n_{s}}$ is as given in Eq.(\ref{deltan}), while $\Delta _{2s}$
and $\Delta _{3s}$ are given by%
\begin{eqnarray}
\Delta _{2s} &\equiv &\frac{T_{\perp s}}{M_{s}}\left( \gamma _{1}+3\gamma
_{3}V_{s}^{2}\right) -\frac{\gamma _{2}2T_{\perp s}^{2}}{BM_{s}}+ \\
&&+\frac{\gamma _{3}T_{\perp s}}{M_{s}^{2}}\left( T_{\parallel s}+4T_{\perp
s}\right) ,  \notag \\
\Delta _{3s} &\equiv &\frac{I\gamma _{2}T_{\perp s}}{RB^{2}M_{s}}\left(
2T_{\perp s}-T_{\parallel s}\right) +  \label{delta3-} \\
&&+\frac{I\left( T_{\parallel s}-T_{\perp s}\right) }{RBM_{s}}\left( \gamma
_{1}+3\gamma _{3}V_{s}^{2}\right) +  \notag \\
&&+\frac{I\gamma _{3}}{RBM_{s}^{2}}\left( 3T_{\parallel s}^{2}-4T_{\perp
s}^{2}+T_{\parallel s}T_{\perp s}\right) .  \notag
\end{eqnarray}%
Finally, in the limit of isotropic temperatures, the solution for the number
density is%
\begin{equation}
\Delta _{n_{s}}\equiv V_{s}\left[ \gamma _{1}+\gamma _{3}\left( \frac{5T_{s}%
}{M_{s}}+V_{s}^{2}\right) \right] ,  \label{deltaniso}
\end{equation}%
where now $\gamma _{2}=0$,%
\begin{eqnarray}
\gamma _{1} &\equiv &\left\{ \frac{cM_{s}R}{Z_{s}e}K+\frac{M_{s}V_{s}}{T_{s}}%
\left[ 1+\psi A_{3s}\right] \right\} , \\
\gamma _{3} &\equiv &\left\{ \frac{cM_{s}^{2}R}{Z_{s}e}\frac{A_{2s}}{2T_{s}}%
\right\} ,
\end{eqnarray}%
and $K\equiv \left[ A_{1s}+A_{2s}\left( \frac{Z_{s}e\Phi _{s}^{eff}-\frac{%
Z_{s}e}{c}\psi \Omega _{s}(\psi )}{T_{s}}-\frac{1}{2}\right) \right] $. For
calculating the flow velocity from Eq.(\ref{deltaphi}), in the same limit, $%
\Delta _{n_{s}}$ is as given by Eq.(\ref{deltaniso}), $\Delta _{2s}$ reduces
to%
\begin{equation}
\Delta _{2s}\equiv \frac{T_{s}}{M_{s}}\left( \gamma _{1}+3\gamma
_{3}V_{s}^{2}\right) +\frac{5\gamma _{3}T_{s}^{2}}{M_{s}^{2}}
\end{equation}%
and $\Delta _{3s}\equiv 0$ since, in Eq.(\ref{delta3-}), the second and
third terms on the right hand side necessarily vanish, while the first one
is proportional to $A_{4s}\equiv \frac{\partial \alpha _{s}}{\partial \psi }%
, $ where $\alpha _{s}\left( \psi \right) \equiv \frac{B}{\Delta _{T_{s}}}%
\equiv 0,$ and hence vanishes too.

\bigskip






\bigskip

\begin{thebibliography}{99}
\bibitem{Frank} J. Frank, A. King and D. Raine, \textit{Accretion power in
astrophysics} (Cambridge University Press, 2002).

\bibitem{Vietri} M. Vietri, \textit{Astrofisica delle alte energie}
(Bollati-Boringhieri 2006, ISBN 88-339-5773-X).

\bibitem{Balbus1998} S.A. Balbus and J.F. Hawley, Rev. Mod. Phys. \textbf{70}%
, 1 (1998).

\bibitem{Sera01} S. Markoff, H. Falcke, F. Yuan, and P.L. Biermann, Astron.
Astrophys. \textbf{379}, L13-L16 (2001).

\bibitem{Melia} F. Melia and H. Falcke, Annu. Rev. Astron. Astrophys. 
\textbf{39}, 309--52 (2001).

\bibitem{Ferraro1937} V.C. Ferraro, Mon. Not. R. Astron. Soc. \textbf{97},
458 (1937).

\bibitem{Mestel1961} L. Mestel, Mon. Not. R. Astron. Soc. \textbf{122}, 473
(1961).

\bibitem{Lovelace1976} R.V.E. Lovelace, Nature \textbf{262}, 649 (1976).

\bibitem{Blanford1976} R.D. Blanford, Mon. Not. R. Astron. Soc. \textbf{176}%
, 465 (1976).

\bibitem{Blanford-Payne1982} R.D. Blanford and D.G. Payne, Mon. Not. R.
Astron. Soc. \textbf{199}, 883 (1982).

\bibitem{Miller2001} E. Szuszkiewicz and J.C. Miller, Mon. Not. R. Astron.
Soc. \textbf{328}, 36-44 (2001).

\bibitem{Coppi1} B. Coppi, Phys. Plasmas \textbf{12}, 057302 (2005).

\bibitem{Coppi2} B. Coppi and F. Rousseau, Astrophys. J. \textbf{641},
458-470 (2006).

\bibitem{Cremasch2008-1} C. Cremaschini, A. Beklemishev, J. Miller and M.
Tessarotto, AIP Conf. Proc. \textbf{1084}, 1067-1072 (2008), arXiv:0806.4522.

\bibitem{Cremasch2008-2} C. Cremaschini, A. Beklemishev, J. Miller and M.
Tessarotto, AIP Conf. Proc. \textbf{1084}, 1073-1078 (2008), arXiv:0806:4923.

\bibitem{Quataert2002} E. Quataert, W. Dorland and G.W. Hammett, Astrophys.
J. \textbf{577}, 524-533 (2002).

\bibitem{Quataert2007} P. Sharma, E. Quataert, G.W. Hammett and J.M. Stone,
Bull. Am. Phys. Soc. \textbf{52}, 11 (2007).

\bibitem{Quataert2007B} P. Sharma, E. Quataert, G.W. Hammett and J.M. Stone,
Astrophys. J. \textbf{667}, 714-723 (2007).

\bibitem{NarayanA} R. Narayan, R. Mahadevan and E. Quataert, 1998 in Theory
of Black Hole Accretion Discs, ed. M. Abramowicz, G. Bjornsson and J.
Pringle, Cambridge University Press, 148.

\bibitem{NarayanB} R. Narayan and I. Yi, Astrophys. J \ \textbf{452}, 710
(1995).

\bibitem{Mahajan01} P. Bhaskaran and V. Krishan, Astrophysics and Space
Science \textbf{232}, 65-78 (1995).

\bibitem{Mahajan02} S.M. Mahajan, Phys. Fluids B \textbf{1}, 143 (1989).

\bibitem{Mahajan03} S.M. Mahajan, Phys. Fluids B \textbf{1}, 2345 (1989).

\bibitem{Catto1987} P.J. Catto, I.B. Bernstein and M. Tessarotto, Phys.
Fluids B \textbf{30}, 2784 (1987).

\bibitem{Balbus1991} S.B. Balbus and J.F. Hawley, Astrophys. J. \textbf{376}%
, 214-222 (1991).

\bibitem{Snyder1997} P.B. Snyder, G.W. Hammett and W. Dorland, Phys. Plasmas 
\textbf{4}, 11 (1997).

\bibitem{Sharma2006} P. Sharma, G.W. Hammett, E. Quataert and J.M. Stone,
Astrophys. J. \textbf{637}, 952-967 (2006).

\bibitem{Catto1977} P.J. Catto and K.T. Tsang, Phys. Fluids \textbf{20}, 396
(1977).

\bibitem{Catto1978} P.J. Catto, Plasma Phys. \textbf{20}, 719 (1978).

\bibitem{Bernstein1985} I.B. Bernstein and P.J. Catto, Phys. Fluids \textbf{%
28}, 1342 (1985).

\bibitem{Bernstein1986} I.B. Bernstein and P.J. Catto, Phys. Fluids \textbf{%
29}, 3897 (1986).

\bibitem{Berk1967} H.L. Berk and A.A. Galeev, Phys. Fluids \textbf{10}, 441
(1967).

\bibitem{Tang1992} A. Artun and W.M. Tang, Phys. Fluids B \textbf{4}, 1102
(1992).

\bibitem{Bondeson1994} A. Bondeson and D.J. Ward, Phys. Rev. Lett. \textbf{72%
}, 2709 (1994).

\bibitem{Hinton1985} F.L. Hinton and S.K. Wong, Phys. Fluids \textbf{28},
3082 (1985).

\bibitem{Tessarotto1992} M. Tessarotto and R.B. White, Phys. Fluids B\textbf{%
\ 4}, 859 (1992).

\bibitem{Tessarotto1993} M. Tessarotto and R.B. White, Phys. Fluids B 
\textbf{5}, 3942 (1993).

\bibitem{Tessarotto1994} M. Tessarotto, J.L. Johnson and L.J. Zheng, Phys.
Plasmas \textbf{2}, 4499 (1995).

\bibitem{Beklemishev1999} A.J. Brizard and A.A. Chan, Phys. Plasmas \textbf{6%
}, 4548 (1999).

\bibitem{Beklemishev2004} A. Beklemishev and M. Tessarotto, Astron.
Astrophys. \textbf{428}, 1 (2004).

\bibitem{Cremaschini2006} M. Tessarotto, C. Cremaschini, P. Nicolini and A.
Beklemishev, Proc. 25th RGD (International Symposium on Rarefied gas
Dynamics, St. Petersburg, Russia, July 21-28, 2006), Ed. M.S. Ivanov and
A.K. Rebrov (Novosibirsk Publ. House of the Siberian Branch of the Russian
Academy of Sciences), p.1001 (2007), ISBN/ISSN: 978-5-7692-0924-6,
arXiv:physics/0611114.

\bibitem{Cremaschini2008} C. Cremaschini, M. Tessarotto, P. Nicolini and A.
Beklemishev, AIP Conf. Proc. \textbf{1084}, 1091-1096 (2008),
arXiv:0806.4663.

\bibitem{Littlejohn1979} R.G. Littlejohn, J. Math. Phys. \textbf{20}, 2445
(1979).

\bibitem{Littlejohn1981} R.G. Littlejohn, Phys. Fluids\textbf{\ 24}, 1730
(1981).

\bibitem{Littlejohn1983} R.G. Littlejohn, J. Plasma Phys.\textbf{\ 29}, 111
(1983).

\bibitem{Dubin1983} D.H.E. Dubin, J.A. Krommes, C. Oberman and W.W. Lee,
Phys.Fluids \textbf{11}, 569 (1983).

\bibitem{Hahm1988} T.S. Hahm, W.W. Lee and A. Brizard, Phys. Fluids \textbf{%
31}, 1940 (1988).

\bibitem{Balescu} B. Weyssow and R. Balescu, J. Plasma Phys. \textbf{35},
449 (1986).

\bibitem{Meiss1990} J.D. Meiss and R.D. Hazeltine, Phys. Fluids B\textbf{\ 2}%
, 2563 (1990).

\bibitem{Kruskal} M. Kruskal, J. Math. Phys. Sci. \textbf{3}, 806 (1962).

\bibitem{meiss} R.D. Hazeltine and J.D. Meiss, \textit{Plasma confinement}
(Addison-Wiley Publishing Company, 1992).
\end{thebibliography}
\end{document}